\documentclass{bfk2}
\usepackage{epsfig}
\usepackage{url}
\usepackage{mathbbol}
\usepackage{booktabs}
\usepackage{fancyvrb}

\newenvironment*{codeblock}%
  {\Verbatim[fontsize=\small,%
             numbers=left,%
             baselinestretch=0.9,%
             xleftmargin=2em,%
             numbersep=1em,
             commandchars=\\\{\}]}
  {\endVerbatim}

\newcommand*{\CC}{C\({}^{{}_{++}}\)}  

\date{29 January 2001}
\title[GiNaC]{Introduction to the GiNaC Framework for Symbolic
  Computation within the \CC\ Programming Language}
\author[C. Bauer, A. Frink, R. Kreckel]%
{Christian Bauer, Alexander Frink, Richard Kreckel \\
Institute of Physics, Johannes-Gutenberg-University, Mainz, Germany}

\begin{document}
\maketitle
\begin{abstract}
  The traditional split-up into a low level language and a high level
  language in the design of computer algebra systems may become
  obsolete with the advent of more versatile computer languages.  We
  describe GiNaC, a special-purpose system that deliberately denies
  the need for such a distinction.  It is entirely written in \CC\ and
  the user can interact with it directly in that language.  It was
  designed to provide efficient handling of multivariate polynomials,
  algebras and special functions that are needed for loop calculations
  in theoretical quantum field theory.  It also bears some potential
  to become a more general purpose symbolic package.
\end{abstract}

\section{Introduction}

Historically, in the design and implementation of symbolic computation
engines, the focus has always been rather on algebraic capabilities
than on language design.  The need for efficient computation in all
fields of science has led to the development of powerful algorithms.
Thus, the border line between inexact numerical and exact analytical
computation has moved, such that more computation may be done exactly
before resorting to numerical methods.  This development has had great
influence on the working practice in such fields as engineering (e.g.
robotics), computer science (e.g.  networks), physics (e.g. high
energy physics) and even mathematics (from group theory to automatic
theorem proving in geometry).

This border line between analytical and numerical methods, however,
has quite often remained a painful obstacle in the successful
application of computer algebra systems (CASs).  Usually, some results
are obtained with the help of a CAS and later these results are
integrated into some other program.  This is not only restricted to
numerical results, where one frequently translates the output of some
CAS to C or even FORTRAN.  It is questionable whether the distinction
between one language for writing down symbolical algebra and one for
obtaining numerical results and maybe a third one for integrating
everything in a graphical user interface has any reason other than
historical ones.  In our experience it frequently leads to confusions;
the \verb|xloops| project~\cite{xloops98} has somewhat suffered from
this.

The chapter ``A Critique of the Mathematical Abilities of CA Systems''
in~\cite{CASguide} has a section called ``Mathematics versus Computer
Science'' where some misbehaviours of common CASs are shown.  There,
the first test tries to find out if a global variable is accessed in
some local context, in particular within sums, products, limits and
integrals.  Of the seven systems tested, only Derive passed the test.
Even explicitly declaring a variable to be local does not always spare
the programmer surprises: MapleV Releases 3 to 5 for instance do not
honor local variables if they are created by concatenating strings
using the operator dot~({\tt .}), a feature people often feel tempted
to use for elegant subscripting.  So created variables may be used as
lvalues in assigning a {\em local} variable within a procedure but the
result is a modified {\em global} variable.\footnote{We have been told
  that Maple6 still suffers from this problem.}  Such violations of
scope have repeatedly led to subtle bugs within \verb|xloops|.  They
are notoriously difficult to disentangle since they go undetected
until some other part of the program breaks.

The general picture is that most currently used CASs are
linguistically rather impoverished and put up high obstacles to the
design of combined symbolical/numerical/graphical programs.  An
incomplete look into the toolchest of a \CC\ developer throws some
light on the features that any professional programmer will miss
from common CA systems:
\begin{itemize}
\item[\(\bullet\)] structured data types like \verb|struct|s and
  \verb|class|es instead of unnamed lists of lists,
\item[\(\bullet\)] the object oriented (OO) programming paradigm in
  general,
\item[\(\bullet\)] {\tt template}s, which allow for generic (i.e.
  type-independent) programming even in a strongly typed language,
\item[\(\bullet\)] the Standard Template Library (STL), which provides
  convenient classes for many kinds of containers with asymptotically
  ideal access methods and to a large degree container-independent
  algorithms ({\tt sort}, etc) to be instantiated by the programmer.
\item[\(\bullet\)] modularization facilities like {\tt namespace}s,
\item[\(\bullet\)] powerful development tools like editors (e.g. with
  automatic indentation and syntax highlighting), debuggers,
  visualization tools, documentation generators, etc.,
\item[\(\bullet\)] flexible error handling with exceptions,
\item[\(\bullet\)] last, but not least: an established standard
  (\citeauthor{ISO14882}) which guards the programmer from arbitrary
  changes made by the manufacturer which break existing code.
\end{itemize}

Solutions for those problems so far are restricted to allowing calls
to CAS functionality from other languages and the already mentioned
code generators.  At most, bridges are built to cross the gap, no
unification of the two worlds is achieved.

\subsection{The goal}

Loop calculations in quantum field theory (QFT) are one example for
such a combined symbolical and numerical effort.  The \(n\)-fold
nested integrals arising there are solved with specialized methods
that demand efficient handling of order \(10^3\)-\(10^6\) symbolic
terms.  At the one-loop level, Feynman graphs can be expressed
completely analytically and so in the early 90's our group started to
build up the program package \verb|xloops| based on Maple.  The
continuation of \verb|xloops| with Maple up to the two-loop level
turned out to be very difficult to accomplish.  There were numerous
technical issues of coding such as the ones outlined above as well as
a nasty restriction built into MapleV of no more than \(2^{16}\) terms
in sums.

An analysis of \verb|xloops| showed, however, that only a small part
of the capabilities of Maple is actually needed: composition of
expressions consisting of symbols and elementary functions,
replacement of symbols by expressions, non-commuting objects,
arbitrarily sized integers and rationals and arbitrary precision
floats, collecting expressions in equal terms, power series expansion,
simplification of rational expressions and solutions of symbolic
linear equation systems.

It is possible to express all this directly in \CC\ if one introduces
some special classes of symbols, sums, products, etc.  More generally,
one wishes to freely pass general expressions to functions and back.
Here is an example of how some of these things are actually expressed
in \CC\ using the GiNaC\footnote{GiNaC is a recursive acronym for {\em
    GiNaC is Not a CAS}.} framework:
\begin{codeblock}
#include <ginac/ginac.h>
using namespace GiNaC;

ex HermitePoly(const symbol & x, int n)
\{
    const ex HGen = exp(-pow(x,2));
    \textit{// uses the identity H_n(x) == (-1)^n exp(x^2) (d/dx)^n exp(-x^2)}
    return normal(pow(-1,n) * HGen.diff(x, n) / HGen);
\}

int main(int argc, char **argv)
\{
    int degree = atoi(argv[1]);
    numeric value = numeric(argv[2]);
    symbol z("z");
    ex H = HermitePoly(z,degree);
    cout << "H_" << degree << "(z) == " 
         << H << endl;
    cout << "H_" << degree << "(" << value << ") == "
         << H.subs(z==value) << endl;
    return 0;
\}
\end{codeblock}
When this program is compiled and called with \(11\) and \(0.8\) as
command line arguments it will readily print out the 11th Hermite
polynomial together with that polynomial evaluated numerically at
\(z=0.8\):
\begin{codeblock}
H_11(z) == -665280*z+2217600*z^3-1774080*z^5+506880*z^7-56320*z^9+2048*z^11
H_11(0.8) == 120773.8855954841959
\end{codeblock}
Alternatively, it may also be called with an exact rational second
argument \(4/5\):
\begin{codeblock}
H_11(z) == -665280*z+2217600*z^3-1774080*z^5+506880*z^7-56320*z^9+2048*z^11
H_11(4/5) == 5897162382592/48828125
\end{codeblock}
It calls the subroutine \verb|HermitePoly| with the symbolic variable
\(z\) and the desired order as arguments.  There, the Hermite
polynomial is computed in a straightforward way using a Rodrigues
representation.  The \verb|normal()| call therein cancels the
generators \verb|HGen| in numerator and denominator.  Note that the
operators \verb|*| and \verb|/| have been overloaded to allow
expressive construction of composite expressions and that both
object-style method invocations ({\em obj}\verb|.f(|{\em arg}\verb|)|)
as well as function-style calls (\verb|f(|{\em obj}\verb|,|{\em
  arg}\verb|)|) are possible.  Technically, the whole GiNaC library is
hidden in a namespace that needs to be imported (for instance with the
\verb|using| directive in line 2) in order to allow easy integration
with other packages without potential name clashes.  This is just a
crude example that invites for obvious refinement like parameter
checking or rearranging the polynomial in order to make it less
sensitive to numerical rounding errors.

Since pattern matching is something that doesn't blend very naturally
into the context of a declarative language like \CC, GiNaC takes care
to use term rewriting systems which bring expressions into equivalent
canonical forms as far as feasible in an economic way.  In addition,
specialized transformations may be invoked by the user, for instance
an \verb|.expand()| method for fully expanding polynomials or a
\verb|.normal()| method for polynomial GCD cancellation.  As for
instance in {\sc Form}~\cite{VermaserenForm} the user alone is
responsible for deciding the order of steps to take in some
application, there are only very few rules built into GiNaC.  The only
kind of pattern matching we want to allow is an atomic one, where
inside an expression a symbol (or a list of symbols) is replaced by
other expressions in this fashion: \verb|(5*a).subs(a==b)|
\(\Rightarrow\) \verb|5*b|.  We believe that such a conservative
restriction should be acceptable to programmers of large systems since
the potential ambiguities introduced by pattern matching and
overlapping rules can be rather subtle.

\section{The implementation}

\begin{figure}['t']
\epsfig{file=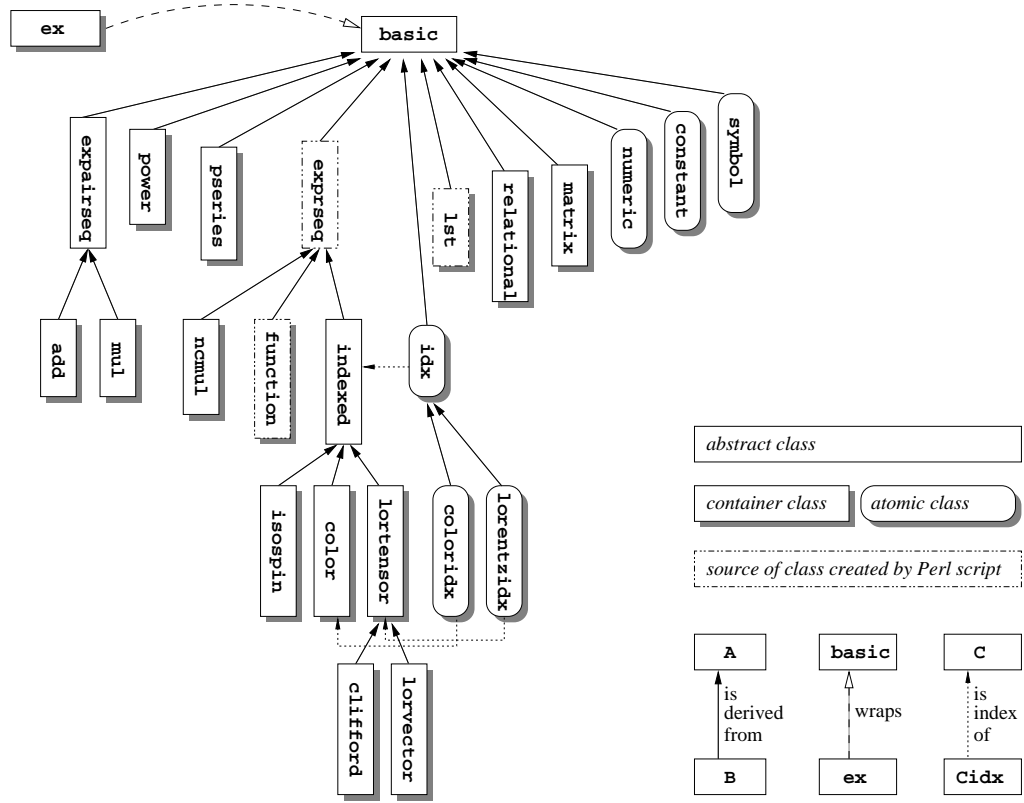,width=\textwidth}
\caption{The GiNaC class hierarchy and some of the relations between the
  classes.}\label{fig:classhierarchy}
\end{figure}

The implementation of GiNaC follows an OO philosophy: all algebraic
classes that may be manipulated by the system are derived from an
abstract base class called \verb|basic|\footnote{Strictly speaking,
  GiNaC does not have any abstract base classes in the \CC\ sense,
  since there are defaults for all methods.  We therefore define an
  abstract base class to be one which does not make sense to
  instantiate.}.  Some of the classes are atomic (symbols,
numbers\dots), others are container classes (sums, products\dots).
Since at run-time container classes must be flexible enough to store
different objects whose size must, however, be defined already at
compile-time, we define the class of all expressions, simply called
\verb|ex|.  It is a wrapper class that stands outside the class
hierarchy and it contains mainly a pointer to some object of the class
hierarchy.  The container classes thus may be restricted to hold
objects of the wrapper class \verb|ex|
(figure~\ref{fig:classhierarchy}).
Because of this ``handle'' character, objects of class \verb|ex| are also
the ones the user creates most of the time.  Most operators have been
overloaded to work within this class and they are the most common
arguments to the functions in GiNaC.  Two obvious drawbacks of this
flexibility are the lack of type-safety at compile-time and possible
performance losses by additional function calls in method invocations.
To some extent, this can be remedied by carefully overloading
specialized functions and operators.  On the other hand, the interplay
between \verb|ex| and \verb|basic| (and all classes derived from it)
enables us to implement an efficient memory management using reference
counting and copy-on-write semantics: multiply occurring expressions
(or subexpressions within an expression tree) are shared in memory and
copied only when they need to be modified in one part of the program.
This happens in a completely transparent way for the user.  In order
to create one's own classes managed this way it suffices to derive them
from class \verb|basic|.
\begin{table}['t']
\begin{tabular*}{\textwidth}{@{}l@{\extracolsep{\fill}}l@{\extracolsep{\fill}}r@{}}
\toprule
class & description & examples \\
\midrule
\verb|symbol| & algebraic symbols & \(x\) \\
\verb|numeric| & polymorphic CLN numbers & 42, \(\frac{7}{3}i\), 0.12345678 \\
\verb|constant| & symbols with associated \verb|numeric| & \(\pi\) \\
\verb|add| & sums of expressions & \(a-2b+3\) \\
\verb|mul| & products of expressions & \(2a^2(x\!+\!y\!+\!z)/b\) \\
\verb|power| & exponentials & \(x^2\), \(a^{(b+c)}\), \(\sqrt{2}\) \\
\verb|pseries| & truncated power series & \(x-\frac{1}{6}x^3+\mathcal{O}(x^5)\) \\
\verb|function| & symbolic functions & \(\sin(2x)\) \\
\verb|lst| & list of expressions & \verb|[|\(x,2y,3+z\)\verb|]| \\
\verb|relational| & relation between two expressions & \(x\)\verb|==|\(y\) \\
\verb|matrix| & matrices (and vectors) of expressions & \scriptsize\((\!\begin{array}{cc}1\!&\!x\\-x\!&\!1\end{array}\!)\) \\
\verb|ncmul| & container for non-commutative objects & \(\gamma_0\gamma_1\)\\
\verb|color|, \verb|coloridx| & \(SU(3)\) Lie algebra element, -index & \(T_a\), \(\delta_{ab}\), \(f_{abc}\) \\
\verb|lortensor|, \verb|lorentzidx| & Lorentz tensor, -index & \(p^\mu, g_{\mu\nu} \) \\
\bottomrule
\end{tabular*}
\caption{List of the most important classes from 
figure~\ref{fig:classhierarchy} and their purpose.}
\label{tab:classoverview}
\end{table}

Table~\ref{tab:classoverview} gives an overview of what classes are
currently provided by GiNaC.  We are now going to describe some of
them.

\subsection{Numbers}

Arbitrarily sized integers, rationals and arbitrary precision floating
point numbers are all stored in the class \verb|numeric|.  This is an
interface that encapsulates the foundation class \verb|cl_N| of Bruno
Haible's \CC\ library {\tt CLN}~\cite{CLN} in a completely transparent
way.

The choice fell to {\tt CLN} because it provides fast and
asymptotically ideal algorithms for all basic operations (Karatsuba
and Sch{\"o}nhage-Strassen multiplication) and a very flexible way of
dealing with rationals and complex numbers. Also, it does not put any
burden of memory management on us since all objects are
reference-counted---just like GiNaC's---so there are no interferences
with garbage collection.  Its polymorphic types are perfectly suited
for implementing a CAS, and indeed were written with this intention in
mind.  For instance, it honors the injection of the naturals into the
rationals and of the complex numbers into the reals: Rationals are
instantanously and efficiently normalized to coprime integer numerator
and denominator and converted to integers if the resulting denominator
is unity and complex numbers are instantanously converted to reals if
the imaginary part vanishes.  Non-exact numbers, i.e. floats and
complex floats are constructed with any user-defined accuracy.

GiNaC provides functions and operators defined on class \verb|numeric|
to the user so the wrapper class \verb|ex| may be circumvented.  This
provides both some level of type-safety as well as a considerable
speedup.

\subsection{Symbols and Constants}

Symbols are represented by objects of class {\tt symbol}.  Thus,
construction of symbols is done by statements like {\tt symbol x,y;}.
In a compiled language like \CC\ the name of a variable is of course
unavailable to the running program.  For printing purposes therefore a
constructor from a string is provided, i.e. {\tt symbol
  x("x"),y("y");}.  This is reminiscent of Common Lisp's~\cite{CLTL}
concept of {\em print name}.  The responsibility for not mixing up
names (as in {\tt symbol x("y"),y("x");}) is entirely laid on the
user.  The string is not at all used for identification of objects.
If omitted, the system will still deal out a unique string.

Unlike in other symbolic system evaluators, expressions may not be
assigned to symbols.  This is a restriction we had to introduce for
the sake of consistency in the non-symbolic language \CC.  It is,
however, possible to substitute a symbol within an expression with
some other expression by calling the {\tt .subs()} method.

Objects of class {\tt constant} behave much like symbols except that
they must return some specific number (if possible to arbitrary
accuracy) when the method {\tt .evalf()} is called.  There are several
predefined constants like \(\pi\) etc. which have an associated
function for numerical evaluation to arbitrary accuracy.  Another
possibility is an associated fixed precision \verb|numeric|.  Thus,
physical constants are easily constructed by the user, as in this
fragment:
\begin{codeblock}
constant qe("qe",numeric(1.60219e-19));  \textit{// elementary charge}
cout << qe << endl;                      \textit{// prints 'qe'}
cout << evalf(qe) << endl;               \textit{// prints '1.60219E-19'}
\end{codeblock}

\subsection{Polynomial arithmetic: the classes {\tt add}, {\tt mul} and {\tt power}}

\begin{figure}['t']
\begin{center}
\epsfig{file=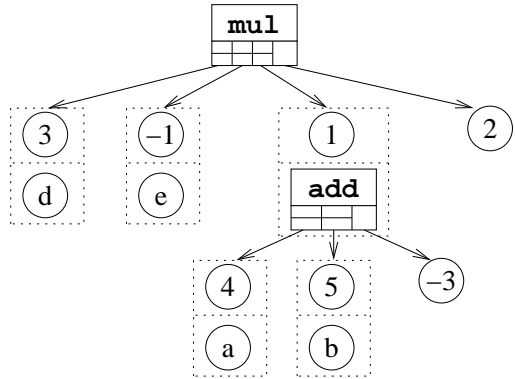,width=.5\textwidth}
\end{center}
\caption{Internal representation of the multivariate rational function $2d^3(4a+5b-3)/e$.}
\label{fig:repmpoly}
\end{figure}

The main object of interest being efficient multivariate polynomials
and rational functions, GiNaC allows the creation of such objects
using the overloaded operators \verb|+|, \verb|-|, \verb|*| and
\verb|/| and the overloaded function
\verb|pow(|\(b\)\verb|,|\(e\)\verb|)| for exponentiation of
expressions \(b\) and \(e\).\footnote{It is also possible to overload
  operator {\tt \char'136} for exponentiation in \CC, but this would
  lead to trouble since it always has lower precedence than {\tt *}.}
When such an object is created, the built-in term rewriting rules of
the classes \verb|add| and \verb|mul| are automatically invoked to
bring it into a canonical form.  Subsequent comparison of such objects
is then easy and further supported by hash values.  Due to the
similarity in the rewriting rules for sums and products the actual
implementation is mostly hidden in class \verb|expairseq|, from which
\verb|add| and \verb|mul| are derived.  The internal representation is
an unexpanded distributive one.  For performance reasons numerical
coefficients in front of monomials in sums and numerical exponents in
products are treated separately (as shown in
figure~\ref{fig:repmpoly}).
The term rewriting rules for class \verb|power| are restricted to
those simplifications that can be done efficiently.

GiNaC provides the usual set of operations on multivariate
polynomials: determination of degree and coefficients, expansion of
products over sums, collection of coefficients of like powers,
conversion of rational expressions to a normal form (where numerator
and denominator are relatively prime polynomials), decomposition of
polynomials into unit part, content, and primitive part, and
polynomial GCD and LCM computation.  For the latter, GiNaC implements
the heuristic polynomial GCD algorithm described
in~\cite{LiaoFatemanGCD}, augmented by additional heuristics such as
cancelling trivial common factors (e.g. \(x^n\)), eliminating
variables that occur only in one polynomial, and special handling of
partially factored polynomials.  If the heuristic algorithm fails,
GiNaC falls back to the subresultant PRS algorithm~\cite{algorithms}.
This approach has so far proved successful for the application in
\verb|xloops|.

\subsection{Power series: the class {\tt pseries}}

Expressions may be differentiated with respect to any symbol and also
expanded as Taylor series or Laurent series.  There is no distinction
between those two.  Series are internally stored in a truncated power
series representation, optionally containing an order term, in a
special class {\tt pseries}.  This class implements efficient
addition, multiplication, and powering (including inversion) of series
and can convert the internal representation to an ordinary GiNaC
expression (polynomial) as well.

A program fragment where the mass increase from special relativity
\(\gamma=(1\!-(\frac{v}{c})^2)^{-1/2}\) is first Taylor expanded and
then inverted and expanded again illuminates the behaviour and syntax
of class \verb|pseries| to some extent:
\begin{codeblock}
symbol v("v"), c("c");
ex gamma = 1/sqrt(1 - pow(v/c,2));
ex gamma_nr = gamma.series(v==0, 6);
cout << pow(gamma_nr,-2) << endl;
cout << pow(gamma_nr,-2).series(v==0, 6) << endl;
\end{codeblock}
Raising the series \(\gamma_{\!{\mbox{\scriptsize nr}}}\) to the power \(-2\) in
line 4 just returns
\(\big(1+\frac{1}{2}(\frac{v}{c})^2+\frac{3}{8}(\frac{v}{c})^4+\mathcal{O}(v^6)\big)^{-2}\).
Only calling the series method again in line 5 makes the output
simplify to \(1-v^2/c^2+\mathcal{O}(v^{6})\).

\subsection{Functions}

\CC\ functions are not suited for symbolic expressions as arguments.
This is so because if the evaluation engine is unable to evaluate the
argument one wishes to return the function itself which would lead to
an infinite recursion.  If \verb|x| is an indeterminate, then
\verb|sin(x)| is supposed to return \verb|sin(x)|.  In order to achive
this behaviour the class \verb|function| is introduced.  Each object
of this class represents a single function (\verb|sin|,
\verb|cos|\dots) and methods for evaluation, differentiation and so on
may be attached to it.  The \CC\ preprocessor is then used to define
wrapper functions that return the corresponding objects of class
\verb|function|.  This allows us to write functions down in \CC\ 
fashion and obtain the behaviour one knows from usual CASs:
\begin{codeblock}
symbol x("x"), y("y");    
ex Do = Pi*(x+y/2);
cout << "sin(" << Do << ") -> " << sin(Do) << endl;
ex Re = Do.subs(y==1);
cout << "sin(" << Re << ") -> " << sin(Re) << endl;
ex Mi = Re.subs(x==11);
cout << "sin(" << Mi << ") -> " << sin(Mi) << endl;
ex Fa = Mi.evalf();
cout << "sin(" << Fa << ") -> " << sin(Fa) << endl;
\end{codeblock}
The above fragment prints:
\begin{codeblock}
sin(Pi*(x+1/2*y)) -> sin(Pi*(x+1/2*y))
sin(Pi*(1/2+x)) -> sin(Pi*(1/2+x))
sin(23/2*Pi) -> -1
sin(36.128315516282622243) -> -1.0
\end{codeblock}
A great many functions are already predefined in GiNaC, some of them,
however, not yet with the full functionality.  For instance, polygamma
functions may not yet be evaluated numerically.

\section{Benchmarks}

Naturally, we want to know how GiNaC performs in comparison with other
systems.  Therefore we subject it and some other symbolic manipulators
to several stress tests on different hardware architectures.  All
tests concentrate on non-\CC\ arithmetics (arbitrary precision instead
of hardware-near \verb|int|, \verb|double|) and symbolic expressions.
GiNaC is superior when it comes to algorithms which largely rely on
machine-near data types.  The first two tests were inspired by typical
operation patterns in elementary particle physics where many different
symbols and deeply nested functions need to be handled.  They are
designed to detect flaws in the memory management and the
implementation of algorithms for manipulation of large container
classes (products, sums\dots).  This is done by having a close look at
the asymptotic runtime behaviour.

\begin{figure}['t']
  \epsfig{figure=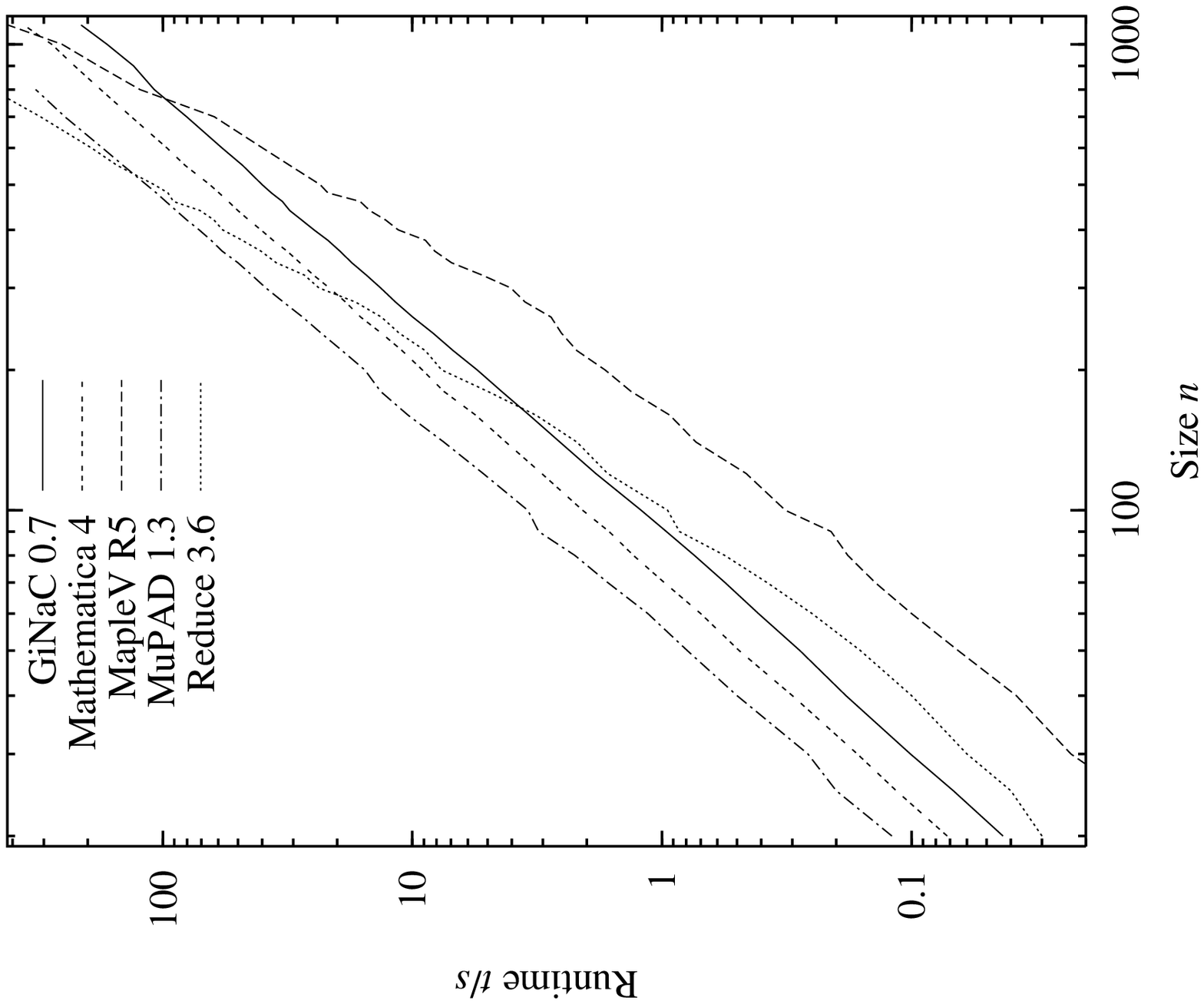,height=.51\textwidth,angle=270}
  \epsfig{figure=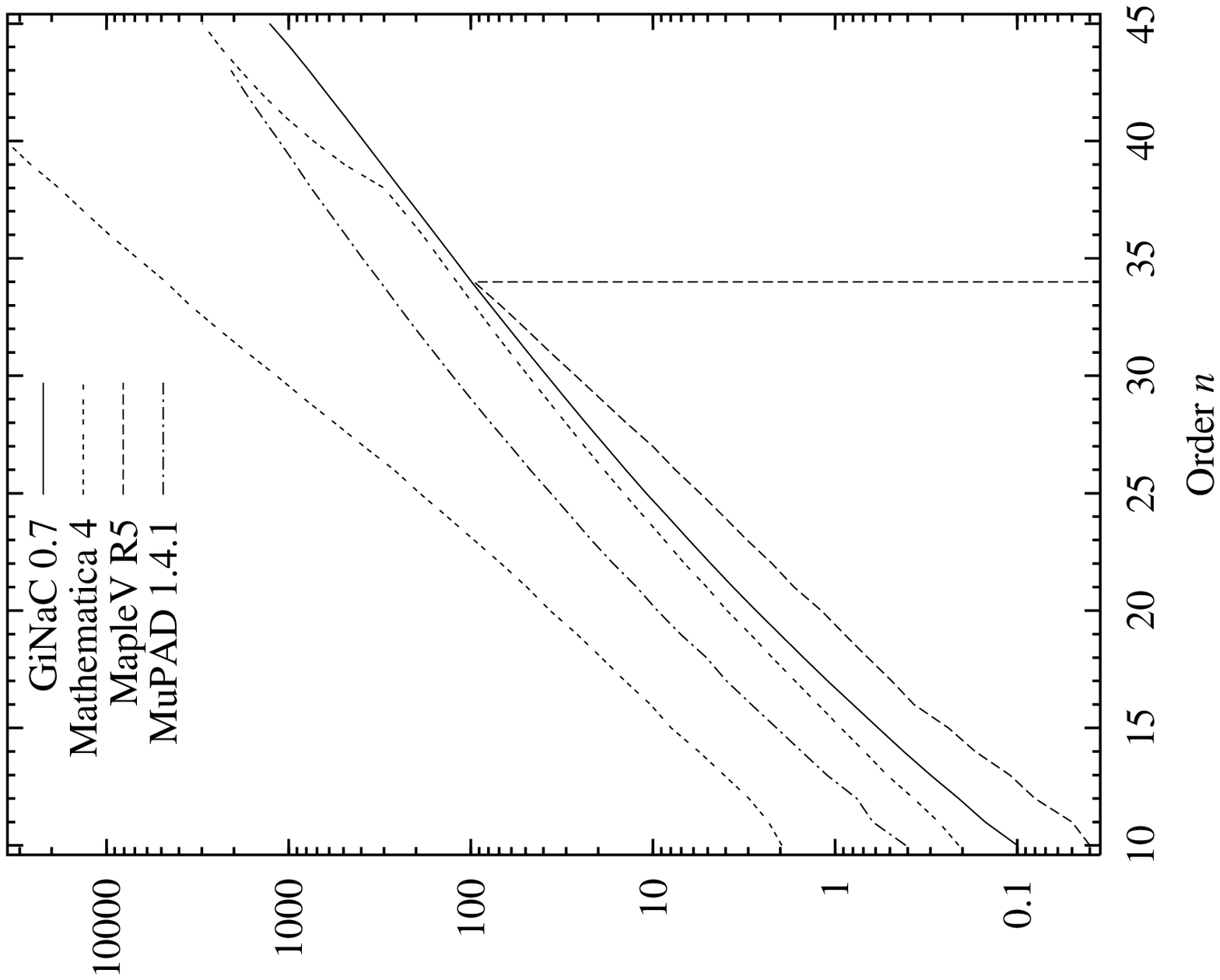,height=.48\textwidth,angle=270}
\caption{Runtimes for a substitute-expand consistency test (left)
  and for series expansion of $\Gamma(x)|_{x=0}$ (right).  The tests
  are described in the text.  The left graph was produced on an
  Alphaserver 8400, the right one on an Intel P-III.}
\label{fig:ginac_gamma}
\end{figure}

The first test (figure~\ref{fig:ginac_gamma}, left) consists of three
steps:
\begin{enumerate}
\item let \(e\) be the expanded sum of \(n\) symbols \(\{a_0,\dots a_{n-1}\}\) 
squared: \(e\leftarrow(\sum_{i=0}^{n-1} a_i)^2\)
\item in \(e\) substitute \(a_0 \leftarrow -\sum_{i=2}^{n-1} a_i\)
\item expand \(e\) again, it collapses to \({a_1}^2\).
\end{enumerate}
The third step is the computationally expensive one.  The system has
to match terms in a sum of \(\approx 2n^2\) elements and eliminate all
but one.  The timings are taken on an Alphaserver 8400 with CPUs of
type EV5 running at 300MHz under Digital Unix 4.  This architecture
was chosen specifically in order to give MapleV a chance, which has an
internal limitation of \(2^{16}-1\) terms in a sum on any other
architecture\footnote{We do not have access to any newer version than
  MapleVR5.  We were informed that the new release Maple6 does not
  suffer from the \(2^{16}-1\) limitation any longer.}.  This turns
out to limit the test to \(n<182\).  This also forced us to resort to
a rather old version of MuPAD because no newer one is available for
the Alpha platform.  The tests were run until we got bored (which we
defined to be 400 seconds).  Further continuation also would have
required more memory since some systems (particularly MuPAD) were
allocating extraordinary amounts of RAM.  The slopes of the curves are
interesting: Those systems that base their memory management on
reference counts exhibit the quadratic scaling one would expect from
the nature of the test while systems with a garbage collector (Maple
and Reducs) start off faster and saturate earlier.

Next, we do a mixed test which besides handling of symbols also
involves handling of large rational numbers and evaluation of
functions at certain points (figure~\ref{fig:ginac_gamma}, right).  We
calculate the expansion of the Gamma function around the pole at
\(x=0\).  The result up to order \(x^2\) is:
\begin{displaymath}
\Gamma(x) = \frac{1}{x} - \gamma 
+ \Big(\frac{\pi^2}{12}\!+\!\frac{\gamma^2}{2}\Big)x 
- \Big(\frac{\pi^2\gamma}{12}\!+\!\frac{\gamma^3}{6}\!+\!\frac{\zeta(3)}{3}\Big)x^2 +\dots
\end{displaymath}
It is not completely clear what other systems are doing internally but
GiNaC's implementation is simple and lacks any optimization.  It falls
back to evaluation of Polygamma functions \(\psi_n(1)\) which in turn
requires the evaluation of Riemann's Zeta function if their argument
is even and, hence, to Bernoulli numbers.  We show two curves for
Mathematica, since this system decides to return the result in the
form of unevaluated Polygamma functions \(\psi_n(1)\).  If one insists
on a result comparable with the other systems one is forced to
introduce calls to \verb|FunctionExpand[]|, which slows the system
down more than an order of magnitude (upper curve).  Without
\verb|FunctionExpand[]| its performance is only slightly worse than
GiNaC's but with a funny excursion at high orders for which we do not
have an explanation.  It should, however, be mentioned that for
Mathematica that ugly result in terms of \(\psi_n(1)\) can under
certain circumstances be acceptable since it may be handled further
without resorting to \(\zeta\)-functions.  This becomes apparent when
one tries to evaluate the coefficients in the resulting series
numerically.  Maple's internal limitation results in the breakdown at
order \(n=35\) in this test.

\begin{table}['t']
\begin{tabular*}{\textwidth}{@{}rlr@{.}lr@{.}lr@{.}lr@{.}lr@{.}l@{}}
\toprule
\multicolumn{2}{@{}l}{}        & \multicolumn{2}{c}{\bf GiNaC} & \multicolumn{2}{c}{\bf MapleV} & \multicolumn{2}{c}{\bf MuPAD} & \multicolumn{2}{c}{\bf Pari-GP} & \multicolumn{2}{c}{\bf Singular} \\
\multicolumn{2}{@{}l}{\bf Benchmark} & \multicolumn{2}{c}{\bf 0.7} & \multicolumn{2}{c}{\bf R5} & \multicolumn{2}{c}{\bf 1.4.1} & \multicolumn{2}{c}{\bf 2.0.19{\boldmath\(\beta\)}} & \multicolumn{2}{c}{\bf 1-3-7} \\
\midrule
A: & divide factorials \(\frac{(1000+i)!}{(900+i)!}\big|_{{}^{i=1}}^{{}_{100}}\) & 0&20  & 6&66   & 1&13   & 0&37  & 19&0 \\
B: & \(\sum_{i=1}^{1000}1/i\)              & 0&019  & 0&08   & 0&10   & 0&041 & 0&54 \\
C: & gcd(big integers)                     & 0&25   & 10&2   & 3&01   & 1&65  & 0&11 \\
D: & \(\sum_{i=1}^{10}iyt^i/(y+it)^i\)     & 0&78   & 0&13   & 1&21   & 0&20  & \multicolumn{2}{l}{NA\footnotemark} \\
E: & \(\sum_{i=1}^{10}iyt^i/(y+|5-i|t)^i\) & 0&63   & 0&05   & 2&33   & 0&11  & \multicolumn{2}{l}{NA\addtocounter{footnote}{-1}\footnotemark} \\
F: & gcd(2-var polys)                      & 0&08   & 0&08   & 0&21   & 0&057 & 0&13 \\
G: & gcd(3-var polys)                      & 2&50   & 2&89   & 3&31   & 99&5  & 0&38 \\
H: & det(rank 80 Hilbert)                  & 10&0   & 33&5   & 42&5   & 3&97  & \multicolumn{2}{l}{CR} \\
I: & invert rank 40 Hilbert                & 3&38   & 6&41   & 12&0   & 0&62  & \multicolumn{2}{l}{CR} \\
J: & check rank 40 Hilbert                 & 1&61   & 2&28   & 2&95   & 0&22  & \multicolumn{2}{l}{UN} \\
K: & invert rank 70 Hilbert                & 22&1   & 92&0   & 74&0   & 5&90  & \multicolumn{2}{l}{CR} \\
L: & check rank 70 Hilbert                 & 9&19   & 21&6   & 14&2   & 1&57  & \multicolumn{2}{l}{UN} \\
M\({}_1\): & rank 26 symbolic sparse, det  & 0&36   & 0&40   & 0&75   & 0&016 & 0&003 \\
M\({}_2\): & rank 101 symbolic sparse, det & 1903&3 & \multicolumn{2}{l}{GU} & \multicolumn{2}{l}{CR} & \multicolumn{2}{l}{CR} & 251&2 \\ 
N: & eval poly at rational functions       & \multicolumn{2}{l}{CR} & \multicolumn{2}{l}{GU} & \multicolumn{2}{l}{CR} & \multicolumn{2}{l}{CR} & \multicolumn{2}{l}{NA} \\
O\({}_1\): & three rank 15 dets (average)  & 43&2   & \multicolumn{2}{l}{GU} & \multicolumn{2}{l}{CR} & \multicolumn{2}{l}{CR} & \multicolumn{2}{l}{CR}   \\
O\({}_2\): & two GCDs                      & \multicolumn{2}{l}{CR} & \multicolumn{2}{l}{UN} & \multicolumn{2}{l}{UN} & \multicolumn{2}{l}{UN} & \multicolumn{2}{l}{UN}   \\
P: & det(rank 101 numeric)                 & 1&10   & 12&6   & 44&3   & 0&09 & 0&85 \\
P': & det(less sparse rank 101)            & 6&07   & 13&3   & 46&2   & 0&38 & 1&25 \\
Q: & charpoly(P)                           & 103&9  & 1429&7  & 741&7   & 0&15 & 4&4  \\
Q': & charpoly(P')                         & 212&8  & 1497&3  & 243&1   & \multicolumn{2}{l}{CR} & 5&0  \\
\bottomrule
\end{tabular*}
\caption{Runtimes in seconds for the tests proposed by Lewis and Wester
(only as far as applicable to GiNaC) on an Intel P-III 450MHz, 384MB RAM 
running under Linux.  Abbreviations used: GU (gave up), CR (crashed, out 
of memory), NA (not available), UN (unable, a prerequisite test failed).}
\label{tab:lwmark}
\end{table}

Next, we apply GiNaC to a number of tests invented by R.~Lewis and
M.~Wester~\cite{LewisWester}.  Strictly speaking, these tests are very
much geared towards the particular capabilities of the system Fermat.
This explains the abundance of benchmarks on Smith and Hermite normal
forms of matrices with numerical entries.  
Nevertheless, we tried to subject GiNaC to these tests where
applicable.\footnote{A fair number of these tests even found their way
  into the suite of GiNaC's regression tests.}  The benchmarks were
rerun on the same machine for those systems available to us and the
rules of the game were slightly simplified: each system was given the
chance to run as long as it needed but it was not allowed to allocate
more than physical memory available.  The tests involving finite
fields and the ones involving Smith and Hermite forms were skipped,
since they are not applicable to GiNaC.  
Tests D and E were slightly rearranged in order to give a meaningful
and comparable result: Maple and MuPAD were forced to cancel common
factors in the result (using \verb|normal|), something Pari-GP does
automatically.
The results shown in
table~\ref{tab:lwmark} are encouraging but show room for optimization.
They also demonstrate some improvement of the other systems (notably
Singular) over the original test performed by Lewis and Wester.  The
reader interested in a detailed description of the tests may
consult~\cite{LewisWester}.

\section{Conclusions and further work}

Although the GiNaC framework was built specifically to become a
symbolic engine for complex computations in quantum field theory it is
our hope that it turns out to be useful for other applications, too.
It provides only modest algebraic knowledge; instead it aims at being
a fast and reliable foundation for combined
symbolical/numerical/graphical projects in \CC.  It may be downloaded
and distributed under the terms of the GNU general public license from
\url{http://www.ginac.de/}.  A tutorial introduction and complete
cross references of the source code can also be found there.
\addtocounter{footnote}{1}\footnotetext{We were informed that
  benchmarks D and E can indeed be performed with Singular --- it is
  just not obvious what the right syntax is.}

Because the cycle edit-compile-execute common for all compiled
languages may be rather tedious during development, care has been
taken in the design of GiNaC to permit an interactive frontend to the
library.  Currently, there are two such interfaces.  The first is the
tiny GiNaC interactive shell \verb|ginsh| for quickly manipulating
some expressions.  It does not provide any programming constructs,
only back-reference to the last printed expressions.  The second is an
interface to the Cint \CC\ interpreter used extensively at CERN in the
object-oriented data analysis framework ROOT~\cite{ROOT}.

Though at this stage GiNaC is already fully functional for the
applications it was originally built for, numerous extensions are
imaginable.  The web page gives some hints in this direction and
further suggestions are more than welcome, as are third-party
contributions.

\subsection*{Acknowledgement}

Part of this work was supported by `Graduiertenkolleg Eichtheorien --
Experimentelle Tests und theoretische Grundlagen' at University of
Mainz.  The authors wish to thank Oliver Welzel for fruitful
discussions in the early phase of the project and to Do Hoang Son for
extensive testing.  Stimulating comments came from Richard Fateman
about efficiency in general, Michael Wester about his benchmarks,
Stephen Watt about practical experiences with memory management
schemes and from Dirk Kreimer and Hubert Spiesberger who contributed
considerably by asking tons of good questions.

\end{document}